%% file: main.tex
\pdfoutput=1

\documentclass[11pt]{article}

\usepackage{acl}  
\input{preamble/packages}

\input{preamble/authors}
\input{preamble/definitions}

\title{The Multilingual Curse at the Retrieval Layer: Evidence from Amharic
}

\begin{document}
\maketitle

\begin{abstract}
Multilingual retrieval increasingly underpins cross-lingual question answering and retrieval-augmented generation. Strong zero-shot scores on multilingual benchmarks are often taken as evidence that current encoders transfer reliably across many languages. We argue that this assumption breaks down for underrepresented, morphologically rich languages, and use Amharic as a diagnostic case. Under a shared passage retrieval protocol covering dense, late-interaction, learned sparse, and cross-encoder paradigms, we compare zero-shot multilingual retrievers, Amharic-fine-tuned multilingual retrievers, and monolingual Amharic retrievers. The strongest zero-shot multilingual retriever underperforms the strongest monolingual Amharic first-stage retriever by 23\% relative MRR@10. Fine-tuning two recent multilingual embedding models on the same Amharic supervision yields 32--60\% relative MRR@10 gains over zero-shot, but the best Amharic-fine-tuned multilingual model remains below the strongest monolingual Amharic retriever. These findings indicate that zero-shot multilingual retrieval is not a sufficient proxy for equitable information access in the LLM era: for underrepresented languages, retrieval must be evaluated and adapted in-language rather than inferred from aggregate multilingual benchmarks.
To foster future research, we publicly release the dataset, codebase, and trained models.\footnote{\url{https://github.com/rasyosef/amharic-neural-ir}}
\end{abstract}

\input{sections/introduction}
\input{sections/Related-work}
\input{sections/Experimental}
\input{sections/Results}
\input{sections/discussion}
\input{sections/Conclusion}
\input{sections/Limitations}
\input{sections/Ethical_considerations}
\input{sections/acknowledgements}


\bibliography{references}

\end{document}

%% file: preamble/packages.tex
\usepackage{times}
\usepackage{latexsym}
\usepackage{multirow}
\usepackage{booktabs}

\usepackage[T1]{fontenc}

\usepackage[utf8]{inputenc}

\usepackage{microtype}

\usepackage{inconsolata}

\usepackage{graphicx}

\usepackage[inline]{enumitem}
\usepackage{xcolor,colortbl}
\usepackage{acronym}
\usepackage{bm}
\usepackage[skip=1mm]{caption}
\usepackage{natbib}
\usepackage{hyperref}
\usepackage{amsmath}
\usepackage{xcolor}  

%% file: preamble/authors.tex
\author{
  Yosef Worku Alemneh\thanks{Equal contribution.} \\
  Independent Researcher \\
  \texttt{yosefwalemneh@gmail.com}
  \And
  Kidist Amde Mekonnen\footnotemark[1] \\
  University of Amsterdam \\
  \texttt{k.a.mekonnen@uva.nl}
  \And
  Maarten de Rijke \\
  University of Amsterdam \\
  \texttt{m.derijke@uva.nl}
}

%% file: preamble/definitions.tex
\definecolor{Gray}{gray}{0.85}
\definecolor{LightCyan}{rgb}{0.88,1,1}

\usepackage[acronym]{glossaries}
\newacronym{OCR}{OCR}{ptical Character Recognition }
\newacronym{IR}{IR}{Information Retrieval}
\newcommand{\heading}[1]{\smallskip\noindent\textbf{#1.}}

\makeatletter
\def\heading{\@startsection{paragraph}{4}{\z@}{.1ex plus 0.1ex minus 0.1ex}{-.25em}{\normalsize\bf}}
\makeatother

\usepackage{xcolor}

\AtBeginDocument{%
  }

\acrodef{DDRO}{direct document relevance optimization}

\hypersetup{
  colorlinks   = true,    
  urlcolor     = blue,    
  linkcolor    = blue,    
  citecolor    = blue      
}

%% file: sections/introduction.tex
\section{Introduction}
\label{sec:intro}

\heading{Multilingual retrieval as an access layer.}
Multilingual retrieval is becoming a central component of how language technologies access information across languages, particularly in semantic search, cross-lingual information access, and retrieval-augmented generation. Recent multilingual embedding models are increasingly positioned as off-the-shelf retrieval encoders for broad language coverage~\citep{embedding_gemma_2025,microsoft2026harrier,wang2024multilingual,zhang-etal-2024-mgte,yu2024arctic}. Strong benchmark scores may therefore make zero-shot multilingual retrieval appear sufficient across languages.

\heading{A retrieval-layer multilingual curse.}
We argue that this reading is premature. The gap is particularly visible, and consequential, for underrepresented, morphologically rich languages. When retrieval representations transfer poorly to such languages, downstream systems that rely on retrieved evidence inherit a retrieval-imposed quality ceiling. As a result, multilingual RAG systems and LLM-based question answering pipelines may degrade in ways that are obscured by aggregate multilingual evaluations.

\heading{Why Amharic.}
We make this concrete using Amharic. Amharic is a Semitic language with over 58 million first- and second-language speakers~\citep{basha2023detection,eberhard2024ethnologue}, written in the Ethiopic (Ge'ez) script. Its root-and-pattern morphology, extensive affixation, and script-specific orthography make it challenging for multilingual encoders, which often rely on subword segmentations and vocabularies that are poorly matched to underrepresented, morphologically rich languages~\citep{rust-etal-2021-good,mekonnen-etal-2025-optimized}.
Amharic is therefore a useful diagnostic case rather than an idiosyncratic one: it is widely used but persistently under-resourced in IR, and the factors that strain retrieval here, namely script, morphology, and limited language-specific supervision, recur across many of the world's languages.

\heading{Benchmark and model suite.}
To stress-test multilingual retrieval transfer, we extend the Amharic passage retrieval benchmark introduced by \citet{mekonnen-etal-2025-optimized} and construct Dataset~V2, with 68K query--passage pairs from AMNEWS~\citep{azime2021amharic}, XL-SUM~\citep{hasan-etal-2021-xl}, Amharic Wikipedia, and AmQA~\citep{abedissa2023amqa}. We evaluate four retrieval paradigms, dense bi-encoders, late-interaction retrievers, learned sparse retrievers, and cross-encoder re-rankers under a shared protocol. We compare zero-shot multilingual retrievers, Amharic-fine-tuned multilingual retrievers, and monolingual Amharic retrievers, including those of \citet{mekonnen-etal-2025-optimized}, with BM25 as a sparse lexical reference. This extends prior Amharic neural IR with two additional retrieval families and a multi-source benchmark.

\heading{Main findings.}
The headline finding is clear: the strongest zero-shot multilingual retriever underperforms the strongest monolingual Amharic first-stage retriever by 23\% relative MRR@10. Fine-tuning two recent multilingual embedding models on the same Amharic supervision narrows the gap substantially, yielding 32--60\% relative MRR@10 gains over their zero-shot counterparts. However, the best Amharic-fine-tuned multilingual model remains below the strongest monolingual Amharic retriever despite having roughly 2.5$\times$ more parameters. Cross-encoder re-ranking on top of an Amharic dense bi-encoder reaches MRR@10 of 0.830, the highest score in our evaluation. 
These results suggest that zero-shot multilingual retrieval is not yet a sufficient proxy for equitable information access in the LLM era. One consequence is that multilingual RAG systems built on zero-shot retrievers may inherit a measurable retrieval-layer quality ceiling for languages like Amharic unless retrieval is evaluated and adapted in-language.

\heading{Contributions.}
\begin{enumerate*}[label=(\roman*)]
\item We use Amharic to test whether strong zero-shot multilingual retrieval transfers to an underrepresented, morphologically rich language.
\item We construct an expanded Amharic passage retrieval benchmark with 68K query--passage pairs and evaluate four retrieval paradigms under a shared protocol.
\item We quantify a persistent zero-shot transfer gap and show that Amharic fine-tuning substantially narrows but does not eliminate it.
\item We release the benchmark, code, evaluation scripts, and model checkpoints to support reproducible research on Amharic retrieval and information access for low-resource languages.\footnote{
\href{https://huggingface.co/datasets/rasyosef/Amharic-Passage-Retrieval-Dataset-V2}{Benchmark};
\href{https://github.com/rasyosef/amharic-neural-ir}{code and evaluation scripts};
\href{https://huggingface.co/collections/rasyosef/amharic-neural-ir-models}{monolingual Amharic models};
\href{https://huggingface.co/collections/kiyam/amharic-fine-tuned-multilingual-retrievers}{Amharic-fine-tuned multilingual models}.}

\end{enumerate*}

%% file: sections/Related-work.tex
\section{Related Work}
\label{sec:related}

\heading{Multilingual retrieval and its evaluation.}
Dense multilingual encoders have become a standard basis for multilingual retrieval, with models such as multilingual E5, Arctic Embed, mGTE, EmbeddingGemma, and harrier-oss-v1 introduced as general-purpose retrieval encoders across languages~\citep{wang2024multilingual,yu2024arctic,zhang-etal-2024-mgte,embedding_gemma_2025,microsoft2026harrier}. Recent multilingual learned sparse retrievers further extend this line to sparse lexical representations by mapping multilingual inputs into a shared lexical space~\citep{nguyen2026milco}. This trend is increasingly consequential for retrieval-augmented generation: multilingual RAG systems rely on retrieval as the evidence-selection layer, and recent work has begun to evaluate retrieval quality in multilingual generation settings~\citep{chirkova-etal-2024-retrieval,thakur-etal-2024-knowing}.

Multilingual retrieval benchmarks have also expanded: MIRACL provides native-annotated ad hoc retrieval data over Wikipedia across 18 languages~\citep{zhang-etal-2023-miracl}, mMARCO extends MS MARCO via machine translation~\citep{bonifacio2021mmarco}, and XOR-TyDi frames open-retrieval QA as cross-lingual retrieval~\citep{asai-etal-2021-xor}. These resources make evaluation more systematic, but coverage remains uneven: underrepresented languages with complex morphology and divergent scripts are less consistently represented, and aggregate scores can obscure language-level degradation. Tokenizer-level analyses further show that multilingual tokenizers often over-segment such languages, producing fragmented representations that can affect downstream modeling~\citep{rust-etal-2021-good}. For retrieval, where representation quality directly shapes ranking, this motivates shared-protocol evaluation at the language level rather than reliance on aggregate multilingual scores.

A common response to weak zero-shot transfer is in-language fine-tuning: adapting multilingual retrievers using language-specific query--passage supervision~\citep{bonifacio2021mmarco,zhang2023toward}. For underrepresented languages, however, such supervision is often weakly labeled, domain-restricted, or limited in scale. For Amharic, prior work has shown that monolingual Amharic retrievers outperform strong multilingual baselines, and that Amharic fine-tuning can substantially improve multilingual retrieval~\citep{mekonnen-etal-2025-optimized}. We build on this evidence by testing whether the same pattern holds under a broader setup: a more diverse Amharic benchmark, additional retrieval paradigms, and a unified comparison of zero-shot multilingual retrievers, Amharic-fine-tuned multilingual retrievers, and monolingual Amharic retrievers. This allows us to quantify the gap that remains after zero-shot transfer and after in-language fine-tuning.

\heading{Amharic neural information retrieval.}
Amharic IR has been constrained by the scarcity of relevance-judged collections. The earliest dedicated resource, 2AIRTC~\citep{yeshambel2020}, is a TREC-style ad hoc retrieval collection, but its limited query set and incomplete judgments make it difficult to use as the sole basis for evaluating neural retrievers. Recent work on Amharic neural IR has therefore relied on weakly supervised passage retrieval benchmarks constructed from public Amharic resources.

The work most directly related to ours is \citep{mekonnen-etal-2025-optimized}, which introduced monolingual Amharic dense retrievers based on Amharic BERT and RoBERTa backbones, evaluated multilingual dense baselines and BM25, fine-tuned a multilingual retriever with Amharic supervision, and trained a ColBERT-style late-interaction retriever. Their benchmark was derived from AMNEWS~\citep{azime2021amharic}, using headlines as queries and article bodies as relevant passages. They showed that monolingual Amharic retrievers outperform strong multilingual embedding baselines, and that Amharic fine-tuning substantially improves multilingual retrieval effectiveness.

We build on this foundation rather than reintroducing Amharic dense retrieval. Our contribution is to extend the prior setup in three directions. First, we evaluate on a multi-source V2 benchmark derived from AMNEWS~\citep{azime2021amharic}, XL-SUM~\citep{hasan-etal-2021-xl}, Amharic Wikipedia, and AmQA~\citep{abedissa2023amqa}, moving beyond a single news-derived source. Second, we broaden the retrieval stack by adding learned sparse retrieval and cross-encoder re-ranking to the dense and late-interaction models studied previously. Third, we provide a shared-protocol comparison of zero-shot multilingual retrievers, Amharic-fine-tuned multilingual retrievers, and monolingual Amharic retrievers. This lets us ask not only whether monolingual Amharic retrieval helps, but also whether recent multilingual encoders close the gap once fine-tuned with the same Amharic supervision.

%% file: sections/Experimental.tex
\section{Experimental Setting}
\label{sec:setup}
We evaluate Amharic passage retrieval using a shared benchmark and protocol that covers monolingual Amharic retrievers, zero-shot multilingual retrievers, Amharic-fine-tuned multilingual retrievers, first-stage retrieval, and second-stage re-ranking. Because the evaluated paradigms differ in supervision format, scoring function, and retrieval backend, we treat the results as benchmark-level comparisons of practical retrieval approaches rather than controlled architectural ablations.

\subsection{Benchmark and evaluation}
\label{sec:task_eval}

\heading{Dataset.}
We use the Amharic Passage Retrieval Dataset~V2, a refined multi-source extension of the benchmark introduced by \citet{mekonnen-etal-2025-optimized}. The dataset is derived from AMNEWS~\citep{azime2021amharic}, XL-SUM~\citep{hasan-etal-2021-xl}, Amharic Wikipedia, and AmQA~\citep{abedissa2023amqa}. Following weakly supervised retrieval practice~\citep{hermann2015teaching,wu2020mind,mekonnen-etal-2025-optimized}, headlines serve as queries and article bodies as relevant passages for news-style sources, while AmQA questions serve as queries and answer-containing paragraphs as relevant passages. After MD5-based deduplication, the benchmark contains 68{,}000 query--passage pairs with a fixed 90/10 train--test split.

\heading{Relevance and metrics.}
Each query has a single labeled positive passage derived from source alignment. Relevance judgments are therefore binary and likely incomplete: unlabeled passages may still be relevant, but only the source-aligned passage is treated as positive. We report Recall@5, Recall@10, MRR@10, and NDCG@10 for first-stage retrieval, and the same metrics at cutoff~10 after re-ranking the top-50 first-stage candidates. Under single-positive supervision, Recall@$k$ should be read as a hit-style measure, while MRR@10 and NDCG@10 reflect rank sensitivity with respect to the aligned positive rather than exhaustive relevance.

\subsection{Retrieval models}
\label{sec:retrieval_models}

\heading{Monolingual Amharic retrievers.}
We use \emph{monolingual Amharic retrievers} to refer to models initialized from Amharic-only backbones and trained for Amharic retrieval, distinguishing them from multilingual encoders evaluated zero-shot or fine-tuned on Amharic supervision. We train monolingual Amharic retrievers across four paradigms: dense bi-encoders~\citep{reimers-gurevych-2019-sentence}, late-interaction retrievers~\citep{khattab2020colbert}, learned sparse retrievers~\citep{formal2021splade}, and cross-encoder re-rankers~\citep{nogueira2019passage}. For each paradigm, we train Medium and Base variants initialized from
\texttt{roberta-medium-amharic} and \texttt{roberta-base-amharic}~\citep{alemneh2024amharicbert}, respectively. Table~\ref{tab:amharic_models} summarizes architectures, training data, and objectives for all eight checkpoints.

\begin{enumerate*}[label=(\roman*)]
\item Dense bi-encoders encode queries and passages independently, mean-pool final hidden states, and $\ell_2$-normalize the resulting embeddings. They are trained with MultipleNegativesRankingLoss~\citep{henderson2017efficient} and Matryoshka representations~\citep{kusupati2022matryoshka}.
\item Late-interaction retrievers use token-level MaxSim scoring and are implemented in PyLate~\citep{PyLate}; they project contextualized token embeddings to 128 dimensions, truncate queries to 32 tokens and passages to 256 tokens, and train with one positive and up to four pre-mined negatives per query.
\item Learned sparse retrievers use SPLADE-style pooling over the tokenizer vocabulary and combine a contrastive ranking objective with FLOPs-based sparsity regularization~\citep{formal2021splade,formal2022distillationhardnegativesampling}.
\item Cross-encoder re-rankers jointly encode query--passage pairs and predict a scalar relevance score using weighted Binary Cross-Entropy with \texttt{pos\_weight}=7.
\end{enumerate*}

\input{tables/amharic_models}
\heading{Multilingual baselines.}
We benchmark five multilingual dense embedding models released for retrieval applications: {multilingual-e5-large-instruct}~\citep{wang2024multilingual}, {snowflake-arctic-embed-l-v2.0}~\citep{yu2024arctic}, {gte-multilingual-base}~\citep{zhang-etal-2024-mgte}, {embeddinggemma-300m}~\citep{embedding_gemma_2025}, and {harrier-oss-v1-270m}~\citep{microsoft2026harrier}. All five are evaluated as {zero-shot multilingual retrievers} using each model's released interface: native tokenizer, pooling strategy, encoding procedure, and model-specific query or passage prompts when recommended by the model authors. For {harrier-oss-v1-270m}, we follow the model card and apply the recommended query-side instruction prompt during zero-shot evaluation, while passages are encoded without an instruction.

Of these multilingual baselines, two recent multilingual embedding models, {embeddinggemma-300m} and {harrier-oss-v1-270m}, are also evaluated as {Amharic-fine-tuned multilingual retrievers}. We fine-tune them on the Dataset~V2 training split using the recipe described in Section~\ref{sec:training_eval}. Holding the training split and adaptation recipe fixed across these baselines reduces confounding from supervision differences and focuses the comparison on multilingual initialization and transfer. We additionally include BM25 as a sparse lexical reference and the monolingual Amharic retrievers from \citet{mekonnen-etal-2025-optimized} as prior in-language reference points.

\subsection{Training and indexing}
\label{sec:training_eval}
All monolingual Amharic models are trained on NVIDIA A100 40GB GPUs with AdamW, mixed precision (FP16), and a fixed random seed of 42. Medium variants are initialized from \emph{roberta-medium-amharic}, and Base variants from \emph{roberta-base-amharic}. Dense bi-encoders and cross-encoders use early stopping on validation NDCG@10; late-interaction and learned sparse models are trained for a fixed number of epochs. Table~\ref{tab:hyperparams} summarizes the per-paradigm training configurations.

For Amharic-fine-tuned multilingual retrievers, we fine-tune {embeddinggemma-300m} and {harrier-oss-v1-270m} on a single NVIDIA H100 GPU using SentenceTransformers with MultipleNegativesRankingLoss wrapped in Matryoshka Loss. Both models are trained for 6 epochs with batch size 32 and 4 gradient accumulation steps, giving an effective batch size of 128. We use a cosine learning-rate schedule with peak learning rate $4\times10^{-5}$, warmup ratio 0.025, and BF16 mixed precision. Training triples pair each query with its aligned positive passage and up to four pre-mined negatives provided in the dataset. For each query, we select the two most similar and two least similar passages from the available negative candidates, using the dataset-provided cosine similarity scores. Fine-tuned multilingual models are evaluated without prompts, matching the prompt-free fine-tuning format.

Tokenization can materially affect retrieval in morphologically rich languages~\citep{rust-etal-2021-good,mekonnen-etal-2025-optimized}; we therefore use the native tokenizer of each backbone and apply a fixed text normalization pipeline across all experiments. Dense bi-encoders use FAISS-based nearest-neighbor retrieval; late-interaction models use the Voyager HNSW index via PyLate; learned sparse retrievers use inverted-index retrieval over sparse term-weight vectors. Retrieval depth and paradigm-specific search settings are held fixed within each comparison.

\begin{table}[t]
\centering
\small
\setlength{\tabcolsep}{1pt}
\renewcommand{\arraystretch}{1.05}
\begin{tabular}{@{}lclcccc@{}}
\toprule
\textbf{Paradigm} & \textbf{LR} & \textbf{Schedule} & \textbf{WU} & \textbf{BS} & \textbf{Epochs} & \textbf{ES} \\
\midrule
Dense       & $6{\times}10^{-5}$ & cosine & 2.5\%    & 64--128 & 6    & Yes \\
Late-inter. & $1{\times}10^{-5}$ & linear & 0\%      & 32      & 4    & --  \\
Sparse      & $6{\times}10^{-5}$ & cosine & 2.5--5\% & 32--48  & 4--6  & --  \\
Cross-enc.  & $4{\times}10^{-5}$ & cosine & 5\%      & 64      & 4    & Yes \\
\bottomrule
\end{tabular}
\caption{
Training hyperparameters for monolingual Amharic retrievers.
LR = learning rate; WU = warmup ratio; BS = batch size; ES = early stopping on validation NDCG@10.
}
\label{tab:hyperparams}
\end{table}

%% file: tables/amharic_models.tex
\begin{table*}[t]
\centering
\setlength{\tabcolsep}{3pt}
\begin{tabular}{@{}lllccrllc@{}}
\toprule
\textbf{Checkpoint} & \textbf{Family} & \textbf{Backbone} & \textbf{Params} & \textbf{Dim.} &
\textbf{Train $N$} & \textbf{Data} & \textbf{Objective} & \textbf{MaxLen} \\
\midrule
\texttt{Embed-Med}     & Dense       & RoBERTa-Med  & \phantom{0}42M & 512              & 122{,}938 & Triples & MNR+MRL    & 510 \\
\texttt{Embed-Base}    & Dense       & RoBERTa-Base & 110M           & 768              & 245{,}876 & Triples & MNR+MRL    & 510 \\
\texttt{ColBERT-Med}   & Late-inter. & RoBERTa-Med  & \phantom{0}42M & 128              & 118{,}938 & Triples & ColBERT-IB & 256 \\
\texttt{ColBERT-Base}  & Late-inter. & RoBERTa-Base & 110M           & 128              & 118{,}938 & Triples & ColBERT-IB & 256 \\
\texttt{SPLADE-Med}    & Sparse      & RoBERTa-Med  & \phantom{0}42M & $\lvert V\rvert$ & 245{,}876 & Triples & SPLADE-SR  & 510 \\
\texttt{SPLADE-Base}   & Sparse      & RoBERTa-Base & 110M           & $\lvert V\rvert$ & 245{,}876 & Triples & SPLADE-SR  & 510 \\
\texttt{Re-rank-Med}   & Cross-enc.  & RoBERTa-Med  & \phantom{0}42M & 1                & 491{,}752 & Pairs   & BCE-w7     & 510 \\
\texttt{Re-rank-Base}  & Cross-enc.  & RoBERTa-Base & 110M           & 1                & 491{,}752 & Pairs   & BCE-w7     & 510 \\
\bottomrule
\end{tabular}
\caption{
Monolingual Amharic retrievers evaluated in this paper.
We include Medium and Base variants across dense, late-interaction, learned sparse, and cross-encoder paradigms; checkpoint names are abbreviated, with full identifiers in the released model collections.
Dim.\ denotes output dimensionality ($\lvert V\rvert$ for sparse vocabulary size), and MaxLen denotes the training truncation length.
Objectives: MNR+MRL = MultipleNegativesRankingLoss with Matryoshka representations; ColBERT-IB = in-batch ColBERT training; SPLADE-SR = sparsity-regularized SPLADE loss; BCE-w7 = binary cross-entropy with \texttt{pos\_weight}=7.
}
\label{tab:amharic_models}
\vspace{-1mm}
\end{table*}

%% file: sections/Results.tex
\vspace{-2mm}
\section{Experimental Results}
\label{sec:results}

We report first-stage retrieval, multilingual fine-tuning, and two-stage re-ranking results on the Amharic Passage Retrieval Dataset~V2. Comparisons follow the shared protocol described in Section~\ref{sec:setup}. Because each query is associated with a single source-aligned positive, Recall@$k$ should be interpreted as a hit-style measure, while MRR@10 and NDCG@10 reflect rank sensitivity with respect to the labeled positive.

\subsection{First-stage retrieval: a persistent zero-shot gap}
\label{sec:firststage}

Table~\ref{tab:first-stage} reports first-stage retrieval effectiveness for BM25, zero-shot multilingual dense retrievers, Amharic-fine-tuned multilingual dense retrievers, monolingual Amharic retrievers from prior work, and the monolingual Amharic retrievers introduced in this work. Three patterns are visible.

\input{tables/first-stage}

\heading{Lexical retrieval remains competitive.}
BM25 reaches MRR@10 of 0.612, outperforming four of the five zero-shot multilingual dense retrievers. Lexical matching is therefore not a weak baseline in this setting: several recent multilingual encoders with hundreds of millions of parameters do not uniformly improve over sparse lexical retrieval on Amharic.

\heading{Zero-shot multilingual retrieval remains below monolingual Amharic retrieval.}
Among zero-shot multilingual dense retrievers, \emph{snowflake-arctic-embed-l-v2.0} performs best, scoring 0.653/0.701 on MRR@10/NDCG@10. The strongest monolingual Amharic first-stage retriever, \emph{ColBERT-Base-Amharic}, reaches 0.803/0.835 with 110M parameters. This corresponds to a 23.0\% relative MRR@10 gap and a 19.1\% relative NDCG@10 gap. The gap is not explained by parameter count: \emph{snowflake-arctic-embed-l-v2.0} has 568M parameters, over five times larger than \emph{ColBERT-Base-Amharic}.

\heading{The gap persists across monolingual Amharic paradigms.}
The strongest monolingual Amharic dense bi-encoder, learned sparse retriever, and late-interaction retriever all outperform the strongest zero-shot multilingual retriever. \emph{Embed-Base-Amharic} reaches MRR@10 of 0.774, \emph{SPLADE-Base-Amharic} reaches 0.754, and \emph{ColBERT-Base-Amharic} reaches 0.803. Late interaction yields the best first-stage effectiveness, while learned sparse retrieval provides an inverted-index alternative that remains competitive with dense retrieval. Within each monolingual Amharic family, Base variants outperform their Medium counterparts, indicating that additional capacity remains useful even in this under-resourced setting.

\subsection{Multilingual fine-tuning narrows but does not close the gap}
\label{sec:ft-multilingual}

The first-stage results show that zero-shot multilingual retrievers underperform monolingual Amharic retrievers. We next ask whether multilingual encoders recover this gap when given the same Amharic supervision. We fine-tune two recent multilingual embedding models, \emph{embeddinggemma-300m} and \emph{harrier-oss-v1-270m}, on the Dataset~V2 training split using the dense bi-encoder fine-tuning recipe described in Section~\ref{sec:setup}. Both fine-tuned models are evaluated without prompts, matching the prompt-free fine-tuning format. For Harrier zero-shot evaluation, we use the recommended query-side instruction prompt; this improves zero-shot NDCG@10 from 0.545 to 0.619, confirming that the prompt is part of the intended retrieval interface.

\noindent \textbf{Fine-tuning produces large gains.}
Fine-tuning substantially improves both multilingual encoders. Gemma improves from MRR@10 0.448 to 0.718, a 60.3\% relative gain. Harrier improves from prompted zero-shot MRR@10 0.576 to 0.760, a 32.0\% relative gain. The same pattern holds for NDCG@10: Gemma improves from 0.489 to 0.753, and Harrier from 0.619 to 0.795. These gains show that multilingual encoders can learn effective Amharic retrieval behavior when given in-language supervision.

\heading{Fine-tuning does not reach the strongest monolingual Amharic retriever.}
Despite these gains, the best Amharic-fine-tuned multilingual model remains below the strongest monolingual Amharic retriever. Harrier fine-tuned reaches 0.760/0.795 on MRR@10/NDCG@10, below \emph{ColBERT-Base-Amharic} at 0.803/0.835. The residual gap is 5.4\% relative MRR@10 and 4.8\% relative NDCG@10. Gemma fine-tuned remains further behind, with MRR@10/NDCG@10 of 0.718/0.753. Thus, in-language fine-tuning narrows the zero-shot gap substantially, but it does not eliminate it.

\heading{Parameter scale is not sufficient.}
The Amharic-fine-tuned multilingual models have 270M--300M parameters, while the strongest monolingual Amharic first-stage retriever has 110M parameters. Harrier fine-tuned is competitive with several monolingual Amharic first-stage models, but it does not match the strongest monolingual Amharic retriever. Gemma fine-tuned also remains below the 42M-parameter \emph{Embed-Medium-Amharic} in MRR@10. These results suggest that multilingual pretraining and parameter scale are not substitutes for language-specific retrieval modeling and supervision.

\subsection{Two-stage re-ranking}
\label{sec:reranking}

We re-rank the top-50 candidates retrieved by \emph{Embed-Base-Amharic} using the two monolingual Amharic cross-encoders. We use the dense bi-encoder as the candidate generator because it provides a standard single-vector first-stage retrieval setup and makes the re-ranking gains directly interpretable. Table~\ref{tab:reranking} shows that re-ranking improves over the dense first-stage retriever in both metrics. \emph{Re-rank-Base-Amharic} raises MRR@10 from 0.774 to 0.830, a 7.2\% relative gain, and NDCG@10 from 0.807 to 0.856, a 6.1\% relative gain. This is the highest score in our evaluation, indicating that joint query--passage scoring helps resolve hard candidate distinctions that are not fully captured by independent bi-encoder representations.

\input{tables/reranking}

%% file: tables/first-stage.tex
\begin{table*}[t]
\centering
\setlength{\tabcolsep}{6pt}
\begin{tabular}{@{}lccccc@{}}
\toprule
\textbf{Model} & \textbf{Params (M)} & \textbf{R@5} & \textbf{R@10} & \textbf{MRR@10} & \textbf{NDCG@10} \\
\midrule
\multicolumn{6}{@{}l}{\textit{Sparse lexical retrieval}} \\
BM25 & -- & 0.734 & 0.789 & 0.612 & 0.655 \\
\midrule
\multicolumn{6}{@{}l}{\textit{Monolingual Amharic retrievers from prior work}} \\
roberta-amharic-text-embedding-medium & \phantom{0}42 & 0.750 & 0.807 & 0.616 & 0.662 \\
roberta-amharic-text-embedding-base   & 110 & 0.790 & 0.844 & 0.657 & 0.703 \\
colbert-roberta-amharic-base          & 110 & 0.860 & 0.899 & 0.736 & 0.776 \\
\midrule
\multicolumn{6}{@{}l}{\textit{Zero-shot multilingual dense retrievers}} \\
embeddinggemma-300m & 300 & 0.558 & 0.621 & 0.448 & 0.489 \\
gte-multilingual-base & 305 & 0.690 & 0.755 & 0.557 & 0.605 \\
harrier-oss-v1-270m & 270 & 0.697 & 0.753 & 0.576 & 0.619 \\
multilingual-e5-large-instruct & 560 & 0.736 & 0.791 & 0.603 & 0.648 \\
snowflake-arctic-embed-l-v2.0 & 568 & 0.795 & 0.848 & 0.653 & 0.701 \\
\midrule
\multicolumn{6}{@{}l}{\textit{Amharic-fine-tuned multilingual dense retrievers}} \\
embeddinggemma-300m + FT & 300 & 0.813 & 0.862 & 0.718 & 0.753 \\
harrier-oss-v1-270m + FT & 270 & 0.860 & 0.903 & 0.760 & 0.795 \\
\midrule
\multicolumn{6}{@{}l}{\textit{Monolingual Amharic retrievers introduced in this work}} \\
SPLADE-Medium-Amharic   & \phantom{0}42 & 0.858 & 0.896 & 0.728 & 0.769 \\
SPLADE-Base-Amharic     & 110 & 0.871 & 0.906 & 0.754 & 0.792 \\
Embed-Medium-Amharic    & \phantom{0}42 & 0.843 & 0.888 & 0.744 & 0.779 \\
Embed-Base-Amharic      & 110 & 0.870 & 0.907 & 0.774 & 0.807 \\
ColBERT-Medium-Amharic  & \phantom{0}42 & \underline{0.882} & \underline{0.913} & \underline{0.778} & \underline{0.811} \\
ColBERT-Base-Amharic & 110 & \textbf{0.902}\rlap{$^{\dagger}$} & \textbf{0.930}\rlap{$^{\dagger}$} & \textbf{0.803}\rlap{$^{\dagger}$} & \textbf{0.835}\rlap{$^{\dagger}$} \\
\bottomrule
\end{tabular}
\caption{
First-stage retrieval results on the Amharic Passage Retrieval Dataset~V2.
Best values are bolded and second-best values are underlined.
Among zero-shot multilingual dense retrievers, \emph{snowflake-arctic-embed-l-v2.0} performs best.
Amharic fine-tuning yields large gains, but the strongest monolingual Amharic retriever achieves the best results across all metrics.
For zero-shot \emph{harrier-oss-v1-270m}, we use the recommended query-side instruction prompt.
$\dagger$ indicates significant improvement over the strongest Amharic-fine-tuned multilingual dense retriever, \emph{harrier-oss-v1-270m + FT}, using paired $t$-tests over per-query scores with $p<0.05$.
}
\label{tab:first-stage}
\end{table*}

%% file: tables/reranking.tex
\begin{table}[t]
\centering
\small
\setlength{\tabcolsep}{1pt}
\begin{tabular}{@{}lcc@{}}
\toprule
\textbf{Model} & \textbf{MRR@10} & \textbf{NDCG@10} \\
\midrule
Embed-Base-Amharic & 0.774 & 0.807 \\
+ Re-rank-Medium-Amharic & 0.805 & 0.835 \\
\textbf{+ Re-rank-Base-Amharic} & \textbf{0.830} & \textbf{0.856} \\
\bottomrule
\end{tabular}
\caption{
Two-stage re-ranking results on the Amharic Passage Retrieval Dataset~V2.
The first row reports the first-stage Embed-Base-Amharic retriever; cross-encoders re-rank its top-50 candidates, with final rank-sensitive metrics computed at cutoff 10.
The Base cross-encoder improves over the first-stage retriever by 7.2\% relative MRR@10 and 6.1\% relative NDCG@10.
}
\label{tab:reranking}
\end{table}

%% file: sections/discussion.tex
\section{Discussion}
\label{sec:discussion}

\heading{Zero-shot scores can hide language-level retrieval failures.}
The results expose a limitation in how multilingual retrieval is often evaluated: aggregate zero-shot performance can mask large language-specific deficits. Amharic is a useful diagnostic case because it combines properties that are common among underrepresented languages but often diluted in multilingual averages: non-Latin script, rich morphology, and limited language-specific retrieval supervision. The strongest zero-shot multilingual retriever reaches 0.653 MRR@10, while the strongest monolingual Amharic first-stage retriever reaches 0.803. This 23\% relative gap occurs in the top-10 region, where retrieval output is typically consumed by users, re-rankers, or downstream generation systems. The consequence is not merely a lower aggregate score; it changes which evidence is made available to later stages of an information access pipeline. This is precisely the failure mode that aggregate multilingual benchmarks can obscure: a model may appear broadly reliable while still providing systematically weaker access for a particular language.

\heading{The gap is not reducible to retrieval architecture.}
The zero-shot gap is not explained by a single modeling choice. Monolingual Amharic dense, learned sparse, and late-interaction retrievers all outperform the strongest zero-shot multilingual baseline. This weakens a purely architectural interpretation: the issue is not simply that one scoring function is stronger than another, but that the multilingual representation space itself transfers imperfectly to Amharic. Architecture still matters, late interaction gives the strongest first-stage retrieval, and cross-encoder re-ranking reaches the highest overall score, but language-specific modeling changes the operating point before architecture-specific gains are considered. In practical terms, a stronger retrieval architecture cannot fully compensate for a representation space that is poorly aligned with the target language. This also helps explain why BM25 remains competitive: lexical overlap, despite its limitations, can be less brittle than a dense representation that fails to align Amharic queries and passages reliably.

\heading{Fine-tuning reveals capacity, but not parity.}
The fine-tuning results complicate a simple failure narrative. Multilingual encoders are not unable to learn Amharic retrieval: Gemma improves from 0.448 to 0.718 MRR@10, and Harrier improves from its prompted zero-shot score of 0.576 to 0.760. This shows that in-language supervision can unlock useful retrieval behavior that is not expressed in the zero-shot setting. However, fine-tuning does not establish parity with monolingual Amharic retrievers. Harrier remains below \emph{ColBERT-Base-Amharic} despite having more than twice as many parameters, and Gemma remains further behind. The relevant comparison is therefore not only whether multilingual encoders improve after fine-tuning, but whether they match in-language alternatives at comparable effectiveness and deployment cost. In our setting, they become competitive, but they do not dominate smaller monolingual Amharic retrievers.

This distinction matters for how adaptation should be interpreted. If fine-tuning only narrowed the gap from a weak zero-shot baseline, the result would support the value of Amharic supervision but would not show that multilingual pretraining is sufficient. The residual gap suggests that the starting representation, tokenizer, and pretraining distribution still matter after supervised adaptation. In-language fine-tuning is therefore necessary for competitiveness, but it is not a substitute for evaluating whether the resulting model matches monolingual alternatives.

\heading{Multilingual RAG evaluation should expose retrieval quality.}
These findings have direct implications for multilingual RAG and LLM-based question answering. Retrieval is the evidence-selection layer: if the relevant passage is not surfaced near the top of the ranked list, the generator receives incomplete or misleading context, and downstream reasoning cannot reliably recover evidence that was never retrieved. For underrepresented languages, RAG evaluation should therefore report retrieval-side quality per language rather than relying only on final generated answers or aggregate multilingual scores. More broadly, the evaluation question should move from whether a model nominally supports a language to whether it retrieves effectively for that language under realistic supervision, architecture, and deployment constraints.

This does not imply that every language requires a separate retrieval stack. Rather, zero-shot multilingual retrieval should not be treated as sufficient evidence of language coverage: for languages such as Amharic, reliable access requires direct retrieval-layer evaluation, fine-tuning tests, and comparison against strong monolingual baselines.

%% file: sections/Conclusion.tex
\vspace{-1mm}
\section{Conclusion}
\label{sec:conclusion}

We use Amharic to test whether strong zero-shot multilingual retrieval transfers to an underrepresented, morphologically rich language. Under a shared passage retrieval protocol, zero-shot multilingual retrievers remain substantially below monolingual Amharic retrievers, with the strongest zero-shot model underperforming the strongest monolingual first-stage retriever by 23\% relative MRR@10. Amharic fine-tuning yields large gains for multilingual encoders, but does not close the gap to the strongest monolingual Amharic retriever. These results show that multilingual retrieval quality cannot be inferred from aggregate zero-shot benchmarks alone: for languages such as Amharic, retrieval must be evaluated and adapted in-language before downstream information-access claims can be trusted.

%% file: sections/Limitations.tex
\vspace{-1mm}
\section{Limitations}
\label{sec:limitations}

Our claims are subject to four main limitations. First, the empirical evidence is limited to Amharic. We argue that the relevant conditions, script divergence, rich morphology, and limited multilingual pretraining coverage also occur in other underrepresented languages, but we do not show that the gap appears at comparable magnitudes elsewhere. Second, Dataset~V2 uses weakly supervised, source-aligned positives: each query has a single labeled relevant passage, while other relevant passages may remain unlabeled. This makes absolute scores conservative and may affect the size of observed gaps relative to a fully judged collection. Third, the multilingual fine-tuning study covers two recent multilingual embedding models, Gemma and Harrier, with zero-shot results for five multilingual baselines; larger or differently instruction-tuned multilingual retrievers may behave differently. 
Finally, our RAG implications are inferred from retrieval metrics rather than measured in an end-to-end generation pipeline. Evaluating whether these retrieval gaps translate into answer-level degradation for Amharic RAG remains future work.

%% file: sections/Ethical_considerations.tex
\section{Ethical Considerations}
\label{sec:ethics}
This work targets Amharic, a widely spoken but under-served language in information access. Dataset~V2 is derived from publicly released sources (AMNEWS, XL-SUM, Amharic Wikipedia, AmQA) under their original licenses; no new human-subject data is collected. Since query--passage labels are weakly derived from source structure, the benchmark is not a human-judged relevance collection. Improved retrieval can aid downstream RAG and QA, but may also propagate source biases if deployed without auditing. We release models and code for reproducible research, not as audited systems for high-stakes deployment.

%% file: sections/acknowledgements.tex
\section*{Acknowledgements}
This work was partially supported by NWO projects EINF-18163/L1, 024.004.022, NWA.\-1389.\-20.\-183, and KICH3.LTP.20.006, and by the European Union's Horizon Europe program under grant agreement No. 101070212. The content reflects only the authors' views.

%% file: references.bib
@String{Computing = "Computing" }

@article{bonifacio2021mmarco,
  title={{mMARCO}: A Multilingual Version of the {MS MARCO} Passage Ranking Dataset},
  author={Bonifacio, Luiz and Jeronymo, Vitor and Abonizio, Hugo Queiroz and Campiotti, Israel and Fadaee, Marzieh and Lotufo, Roberto and Nogueira, Rodrigo},
  journal={arXiv preprint arXiv:2108.13897},
  year={2021}
}

@article{zhang-etal-2023-miracl,
    title = "{MIRACL}: A Multilingual Retrieval Dataset Covering 18 Diverse Languages",
    author = "Zhang, Xinyu  and
      Thakur, Nandan  and
      Ogundepo, Odunayo  and
      Kamalloo, Ehsan  and
      Alfonso-Hermelo, David  and
      Li, Xiaoguang  and
      Liu, Qun  and
      Rezagholizadeh, Mehdi  and
      Lin, Jimmy",
    journal = "Transactions of the Association for Computational Linguistics",
    volume = "11",
    year = "2023",
    address = "Cambridge, MA",
    publisher = "MIT Press",
    url = "https://aclanthology.org/2023.tacl-1.63/",
    doi = "10.1162/tacl_a_00595",
    pages = "1114--1131"
}

@inproceedings{rust-etal-2021-good,
    title = "How Good is Your Tokenizer? {On} the Monolingual Performance of Multilingual Language Models",
    author = "Rust, Phillip  and
      Pfeiffer, Jonas  and
      Vuli{\'c}, Ivan  and
      Ruder, Sebastian  and
      Gurevych, Iryna",
    editor = "Zong, Chengqing  and
      Xia, Fei  and
      Li, Wenjie  and
      Navigli, Roberto",
    booktitle = "Proceedings of the 59th Annual Meeting of the Association for Computational Linguistics and the 11th International Joint Conference on Natural Language Processing (Volume 1: Long Papers)",
    month = aug,
    year = "2021",
    address = "Online",
    publisher = "Association for Computational Linguistics",
    url = "https://aclanthology.org/2021.acl-long.243/",
    doi = "10.18653/v1/2021.acl-long.243",
    pages = "3118--3135"
}

@inproceedings{khattab2020colbert,
author = {Khattab, Omar and Zaharia, Matei},
title = {ColBERT: Efficient and Effective Passage Search via Contextualized Late Interaction over BERT},
year = {2020},
isbn = {9781450380164},
publisher = {Association for Computing Machinery},
address = {New York, NY, USA},
url = {https://doi.org/10.1145/3397271.3401075},
doi = {10.1145/3397271.3401075},
booktitle = {Proceedings of the 43rd International ACM SIGIR Conference on Research and Development in Information Retrieval},
pages = {39–48},
numpages = {10},
location = {Virtual Event, China},
series = {SIGIR '20}
}

@inproceedings{formal2021splade,
author = {Formal, Thibault and Piwowarski, Benjamin and Clinchant, St\'{e}phane},
title = {{SPLADE}: Sparse Lexical and Expansion Model for First Stage Ranking},
year = {2021},
isbn = {9781450380379},
publisher = {Association for Computing Machinery},
address = {New York, NY, USA},
url = {https://doi.org/10.1145/3404835.3463098},
doi = {10.1145/3404835.3463098},
booktitle = {Proceedings of the 44th International ACM SIGIR Conference on Research and Development in Information Retrieval},
pages = {2288–2292},
numpages = {5},
location = {Virtual Event, Canada},
series = {SIGIR '21}
}

@article{nogueira2019passage,
  title={Passage Re-ranking with {BERT}},
  author={Nogueira, Rodrigo and Cho, Kyunghyun},
  journal={arXiv preprint arXiv:1901.04085},
  year={2019}
}

@article{abedissa2023amqa,
  title={{AmQA}: Amharic Question Answering Dataset},
  author={Abedissa, Tilahun and Usbeck, Ricardo and Assabie, Yaregal},
  journal={arXiv preprint arXiv:2303.03290},
  year={2023}
}

@inproceedings{yeshambel2020,
author = {Yeshambel, Tilahun and Mothe, Josiane and Assabie, Yaregal},
title = {{2AIRTC}: The {Amharic} Adhoc Information Retrieval Test Collection},
year = {2020},
isbn = {978-3-030-58218-0},
publisher = {Springer-Verlag},
address = {Berlin, Heidelberg},
url = {https://doi.org/10.1007/978-3-030-58219-7_5},
doi = {10.1007/978-3-030-58219-7_5},
booktitle = {Experimental IR Meets Multilinguality, Multimodality, and Interaction: 11th International Conference of the CLEF Association, CLEF 2020, Thessaloniki, Greece, September 22–25, 2020, Proceedings},
pages = {55–66},
numpages = {12},
location = {Thessaloniki, Greece}
}

@article{azime2021amharic,
  title={An {Amharic} News Text Classification Dataset},
  author={Azime, Israel Abebe and Mohammed, Nebil},
  journal={arXiv preprint arXiv:2103.05639},
  year={2021}
}

@inproceedings{mekonnen-etal-2025-optimized,
    title = "Optimized Text Embedding Models and Benchmarks for {A}mharic Passage Retrieval",
    author = "Mekonnen, Kidist Amde  and
      Alemneh, Yosef Worku  and
      de Rijke, Maarten",
    editor = "Che, Wanxiang  and
      Nabende, Joyce  and
      Shutova, Ekaterina  and
      Pilehvar, Mohammad Taher",
    booktitle = "Findings of the Association for Computational Linguistics: ACL 2025",
    month = jul,
    year = "2025",
    address = "Vienna, Austria",
    publisher = "Association for Computational Linguistics",
    url = "https://aclanthology.org/2025.findings-acl.543/",
    doi = "10.18653/v1/2025.findings-acl.543",
    pages = "10428--10445",
    ISBN = "979-8-89176-256-5"
}

@inproceedings{reimers-gurevych-2019-sentence,
    title = "Sentence-{BERT}: Sentence Embeddings using {S}iamese {BERT}-Networks",
    author = "Reimers, Nils  and
      Gurevych, Iryna",
    editor = "Inui, Kentaro  and
      Jiang, Jing  and
      Ng, Vincent  and
      Wan, Xiaojun",
    booktitle = "Proceedings of the 2019 Conference on Empirical Methods in Natural Language Processing and the 9th International Joint Conference on Natural Language Processing (EMNLP-IJCNLP)",
    month = nov,
    year = "2019",
    address = "Hong Kong, China",
    publisher = "Association for Computational Linguistics",
    url = "https://aclanthology.org/D19-1410/",
    doi = "10.18653/v1/D19-1410",
    pages = "3982--3992"
}

@inproceedings{PyLate,
  author       = {Antoine Chaffin and
                  Rapha{\"{e}}l Sourty},
  editor       = {Meeyoung Cha and
                  Chanyoung Park and
                  Noseong Park and
                  Carl Yang and
                  Senjuti Basu Roy and
                  Jessie Li and
                  Jaap Kamps and
                  Kijung Shin and
                  Bryan Hooi and
                  Lifang He},
  title        = {{PyLate}: Flexible Training and Retrieval for Late Interaction Models},
  booktitle    = {Proceedings of the 34th {ACM} International Conference on Information
                  and Knowledge Management, {CIKM} 2025, Seoul, Republic of Korea, November
                  10-14, 2025},
  pages        = {6334--6339},
  publisher    = {{ACM}},
  year         = {2025},
  url          = {https://github.com/lightonai/pylate},
  doi          = {10.1145/3746252.3761608},
}

@inproceedings{formal2022distillationhardnegativesampling,
author = {Formal, Thibault and Lassance, Carlos and Piwowarski, Benjamin and Clinchant, St\'{e}phane},
title = {From Distillation to Hard Negative Sampling: Making Sparse Neural {IR} Models More Effective},
year = {2022},
isbn = {9781450387323},
publisher = {Association for Computing Machinery},
address = {New York, NY, USA},
url = {https://doi.org/10.1145/3477495.3531857},
doi = {10.1145/3477495.3531857},
booktitle = {Proceedings of the 45th International ACM SIGIR Conference on Research and Development in Information Retrieval},
pages = {2353–2359},
numpages = {7},
location = {Madrid, Spain},
series = {SIGIR '22}
}

@article{henderson2017efficient,
  title={Efficient Natural Language Response Suggestion for Smart Reply},
  author={Henderson, Matthew and Al-Rfou, Rami and Strope, Brian and Sung, Yun-Hsuan and Luk{\'a}cs, L{\'a}szl{\'o} and Guo, Ruiqi and Kumar, Sanjiv and Miklos, Balint and Kurzweil, Ray},
  journal={arXiv preprint arXiv:1705.00652},
  year={2017}
}

@inproceedings{wu2020mind,
    title = "{MIND}: A Large-scale Dataset for News Recommendation",
    author = "Wu, Fangzhao  and
      Qiao, Ying  and
      Chen, Jiun-Hung  and
      Wu, Chuhan  and
      Qi, Tao  and
      Lian, Jianxun  and
      Liu, Danyang  and
      Xie, Xing  and
      Gao, Jianfeng  and
      Wu, Winnie  and
      Zhou, Ming",
    editor = "Jurafsky, Dan  and
      Chai, Joyce  and
      Schluter, Natalie  and
      Tetreault, Joel",
    booktitle = "Proceedings of the 58th Annual Meeting of the Association for Computational Linguistics",
    month = jul,
    year = "2020",
    address = "Online",
    publisher = "Association for Computational Linguistics",
    url = "https://aclanthology.org/2020.acl-main.331/",
    doi = "10.18653/v1/2020.acl-main.331",
    pages = "3597--3606"
}

@inproceedings{hermann2015teaching,
author = {Hermann, Karl Moritz and Ko\v{c}isk\'{y}, Tom\'{a}\v{s} and Grefenstette, Edward and Espeholt, Lasse and Kay, Will and Suleyman, Mustafa and Blunsom, Phil},
title = {Teaching Machines to Read and Comprehend},
year = {2015},
publisher = {MIT Press},
address = {Cambridge, MA, USA},
booktitle = {Proceedings of the 29th International Conference on Neural Information Processing Systems - Volume 1},
pages = {1693–1701},
numpages = {9},
location = {Montreal, Canada},
series = {NIPS'15}
}

@article{wang2024multilingual,
  title={Multilingual {E5} Text Embeddings: A Technical Report},
  author={Wang, Liang and Yang, Nan and Huang, Xiaolong and Yang, Linjun and Majumder, Rangan and Wei, Furu},
  journal={arXiv preprint arXiv:2402.05672},
  year={2024}
}

@article{yu2024arctic,
  title={{Arctic-Embed 2.0}: Multilingual Retrieval Without Compromise},
  author={Yu, Puxuan and Merrick, Luke and Nuti, Gaurav and Campos, Daniel},
  journal={arXiv preprint arXiv:2412.04506},
  year={2024}
}

@inproceedings{kusupati2022matryoshka,
author = {Kusupati, Aditya and Bhatt, Gantavya and Rege, Aniket and Wallingford, Matthew and Sinha, Aditya and Ramanujan, Vivek and Howard-Snyder, William and Chen, Kaifeng and Kakade, Sham and Jain, Prateek and Farhadi, Ali},
title = {Matryoshka Representation Learning},
year = {2022},
isbn = {9781713871088},
publisher = {Curran Associates Inc.},
address = {Red Hook, NY, USA},
booktitle = {Proceedings of the 36th International Conference on Neural Information Processing Systems},
articleno = {2192},
numpages = {17},
location = {New Orleans, LA, USA},
series = {NIPS '22}
}

@inproceedings{basha2023detection,
  title={Detection and Comparative Analysis of Handwritten Words of {Amharic} Language to English Using {CNN}-based Frameworks},
  author={Basha, Shaik Johny and Veeraiah, Duggineni and Charan, Boddu Venkat and Yeddu, Wiltrud Sahithi Joyce and Babu, Devalla Ganesh},
  booktitle={2023 International Conference on Inventive Computation Technologies (ICICT)},
  pages={422--427},
  year={2023},
  organization={IEEE}
}

@book{eberhard2024ethnologue,
  title={Ethnologue: Languages of the World},
  author={Eberhard, David M. and Simons, Gary F. and Fennig, Charles D.},
  edition={27th},
  year={2024},
  publisher={SIL International},
  address={Dallas, Texas},
  url={https://www.ethnologue.com}
}

@inproceedings{hasan-etal-2021-xl,
    title = {{XL}-Sum: Large-Scale Multilingual Abstractive Summarization for 44 Languages},
    author = {Hasan, Tahmid  and
      Bhattacharjee, Abhik  and
      Islam, Md. Saiful  and
      Mubasshir, Kazi  and
      Li, Yuan-Fang  and
      Kang, Yong-Bin  and
      Rahman, M. Sohel  and
      Shahriyar, Rifat},
    booktitle = {Findings of the Association for Computational Linguistics: ACL-IJCNLP 2021},
    month = aug,
    year = {2021},
    publisher = {ACL},
    url = "https://aclanthology.org/2021.findings-acl.413/",
    doi = "10.18653/v1/2021.findings-acl.413",
    pages = "4693--4703"
}

@article{embedding_gemma_2025,
  title={{EmbeddingGemma}: Powerful and lightweight text representations},
  author={Vera, Henrique Schechter and Dua, Sahil and Zhang, Biao and Salz, Daniel and Mullins, Ryan and Panyam, Sindhu Raghuram and Smoot, Sara and Naim, Iftekhar and Zou, Joe and Chen, Feiyang and others},
  journal={arXiv preprint arXiv:2509.20354},
  year={2025}
}

@article{zhang2023toward,
author = {Zhang, Xinyu and Ogueji, Kelechi and Ma, Xueguang and Lin, Jimmy},
title = {Toward Best Practices for Training Multilingual Dense Retrieval Models},
year = {2023},
issue_date = {March 2024},
publisher = {Association for Computing Machinery},
address = {New York, NY, USA},
volume = {42},
number = {2},
issn = {1046-8188},
url = {https://doi.org/10.1145/3613447},
doi = {10.1145/3613447},
journal = {ACM Trans. Inf. Syst.},
month = sep,
articleno = {39},
numpages = {33}
}

@inproceedings{zhang-etal-2024-mgte,
    title = "{mGTE}: Generalized Long-Context Text Representation and Reranking Models for Multilingual Text Retrieval",
    author = "Zhang, Xin  and
      Zhang, Yanzhao  and
      Long, Dingkun  and
      Xie, Wen  and
      Dai, Ziqi  and
      Tang, Jialong  and
      Lin, Huan  and
      Yang, Baosong  and
      Xie, Pengjun  and
      Huang, Fei  and
      Zhang, Meishan  and
      Li, Wenjie  and
      Zhang, Min",
    editor = "Dernoncourt, Franck  and
      Preo{\c{t}}iuc-Pietro, Daniel  and
      Shimorina, Anastasia",
    booktitle = "Proceedings of the 2024 Conference on Empirical Methods in Natural Language Processing: Industry Track",
    month = nov,
    year = "2024",
    address = "Miami, Florida, US",
    publisher = "Association for Computational Linguistics",
    url = "https://aclanthology.org/2024.emnlp-industry.103/",
    doi = "10.18653/v1/2024.emnlp-industry.103",
    pages = "1393--1412"
}

@inproceedings{chirkova-etal-2024-retrieval,
    title = "Retrieval-augmented generation in multilingual settings",
    author = "Chirkova, Nadezhda  and
      Rau, David  and
      D{\'e}jean, Herv{\'e}  and
      Formal, Thibault  and
      Clinchant, St{\'e}phane  and
      Nikoulina, Vassilina",
    editor = "Li, Sha  and
      Li, Manling  and
      Zhang, Michael JQ  and
      Choi, Eunsol  and
      Geva, Mor  and
      Hase, Peter  and
      Ji, Heng",
    booktitle = "Proceedings of the 1st Workshop on Towards Knowledgeable Language Models (KnowLLM 2024)",
    month = aug,
    year = "2024",
    address = "Bangkok, Thailand",
    publisher = "Association for Computational Linguistics",
    url = "https://aclanthology.org/2024.knowllm-1.15/",
    doi = "10.18653/v1/2024.knowllm-1.15",
    pages = "177--188"
}

@inproceedings{asai-etal-2021-xor,
    title = "{XOR} {QA}: Cross-lingual Open-Retrieval Question Answering",
    author = "Asai, Akari  and
      Kasai, Jungo  and
      Clark, Jonathan  and
      Lee, Kenton  and
      Choi, Eunsol  and
      Hajishirzi, Hannaneh",
    editor = "Toutanova, Kristina  and
      Rumshisky, Anna  and
      Zettlemoyer, Luke  and
      Hakkani-Tur, Dilek  and
      Beltagy, Iz  and
      Bethard, Steven  and
      Cotterell, Ryan  and
      Chakraborty, Tanmoy  and
      Zhou, Yichao",
    booktitle = "Proceedings of the 2021 Conference of the North American Chapter of the Association for Computational Linguistics: Human Language Technologies",
    month = jun,
    year = "2021",
    address = "Online",
    publisher = "Association for Computational Linguistics",
    url = "https://aclanthology.org/2021.naacl-main.46/",
    doi = "10.18653/v1/2021.naacl-main.46",
    pages = "547--564"
}

@inproceedings{thakur-etal-2024-knowing,
    title = "``{Knowing} When You Don{'}t Know'': A Multilingual Relevance Assessment Dataset for Robust Retrieval-Augmented Generation",
    author = "Thakur, Nandan  and
      Bonifacio, Luiz  and
      Zhang, Crystina  and
      Ogundepo, Odunayo  and
      Kamalloo, Ehsan  and
      Alfonso-Hermelo, David  and
      Li, Xiaoguang  and
      Liu, Qun  and
      Chen, Boxing  and
      Rezagholizadeh, Mehdi  and
      Lin, Jimmy",
    editor = "Al-Onaizan, Yaser  and
      Bansal, Mohit  and
      Chen, Yun-Nung",
    booktitle = "Findings of the Association for Computational Linguistics: EMNLP 2024",
    month = nov,
    year = "2024",
    address = "Miami, Florida, USA",
    publisher = "Association for Computational Linguistics",
    url = "https://aclanthology.org/2024.findings-emnlp.730/",
    doi = "10.18653/v1/2024.findings-emnlp.730",
    pages = "12508--12526"
}

@misc{microsoft2026harrier,
  author       = {{Microsoft}},
  title        = {\href{https://huggingface.co/microsoft/harrier-oss-v1-270m}{Harrier-OSS-v1-270M}},
  year         = {2026},
  note         = {Hugging Face model card}
}

@inproceedings{nguyen2026milco,
  title={MILCO: Learned Sparse Retrieval Across Languages via a Multilingual Connector},
  author={Nguyen, Thong and Lei, Yibin and Ju, Jia-Huei and Yang, Eugene and Yates, Andrew},
  booktitle={International Conference on Learning Representations},
  year={2026}
}

@misc{alemneh2024amharicbert,
  author       = {Alemneh, Yosef Worku},
  title        = {Amharic BERT and RoBERTa Models},
  year         = {2024},
  howpublished = {\href{https://huggingface.co/collections/rasyosef/amharic-bert-and-roberta}{Hugging Face model collection}}
}
